\documentclass[12pt,a4paper,sumlimits,intlimits, fleqn]{iopart}

\usepackage{fullpage}
\usepackage{framed}
\usepackage{hyperref}
\usepackage{marginnote}
\usepackage{mathrsfs}
\newcommand{\eqref}[1]{(\ref{#1})}

\newcommand{\Cdudu}[4]{C_{#1 \phantom{#2} #3}^{\phantom{#1} #2 \phantom{#3} #4}}
\newcommand{\Tud}[3]{#1^{#2}_{\phantom{#2}#3}}
\def\Lie{\pounds}
\def\nspace{\hspace{-0.02cm}}
\def\eps{\varepsilon}

\def\ii{{}\mathrm{i}{}}

\def\FF{\mathcal{F}}
\def\pd{\partial}
\def\Fdual{{}^\ast \!F}
\def\muG{\mu_{\mathrm{G}}}
\def\PG{P_{\mathrm{G}}}
\def\ie{\emph{i.e.\ }}
\def\eg{\emph{e.g.\ }}
\def\dd{\mathrm{d}}
\def\CC{\mathcal{C}}
\def\DD{\mathcal{D}}
\newcommand{\Cart}[1]{ #1_{\mathrm{cart}}}
\newcommand{\Grav}[1]{ #1_{\mathrm{G}}}
\newcommand{\Dia}{\tilde{\Theta}}
\newcommand{\dfrac}[2]{{\displaystyle \frac{#1}{#2}}}
\newcommand{\KDelta}{\Delta_{\mathrm{K}}}

\usepackage{bm}

\begin{document}
\title{A gravitational energy-momentum and the thermodynamic description of gravity}
\author{G.\ Acquaviva and D.\ Kofro\v{n} and M.\ Scholtz}
\address{Institute of Theoretical Physics, Faculty of Mathematics and Physics, Charles University, V Hole\v{s}ovi\v{c}k\'ach 2, 180~00 Prague, Czech Republic}
\eads{\mailto{gioacqua@utf.troja.mff.cuni.cz}, \mailto{d.kofron@gmail.com}, \mailto{scholtz@utf.mff.cuni.cz}}

\begin{abstract}
A proposal for the gravitational energy-momentum tensor, known in the literature as the \emph{square root of Bel-Robinson tensor}, is analyzed in detail.  Being constructed exclusively from the Weyl part of the Riemann tensor, such tensor encapsulates the geometric properties of free gravitational fields in terms of optical scalars of null congruences: making use of the general decomposition of any energy-momentum tensor, we explore the thermodynamic interpretation of such geometric quantities.  While the matter energy-momentum is identically conserved due to Einstein's field equations, the SQBR is not necessarily conserved and dissipative terms could arise in its vacuum continuity equation.  We discuss the possible physical interpretations of such mathematical properties.
\end{abstract}

\section{Introduction}

The search for a meaningful definition of energy-momentum tensor (EMT) for the gravitational field has been a concern since soon after the formulation of General Relativity \cite{einstein1923hamiltonsches,einstein1918energiesatz,dirac2016general}.  It became clear pretty soon that the equivalence principle itself, a cornerstone of the geometric description of gravity, renders the definition of a local gravitational energy-momentum ambiguous.  In fact, in accordance with the principle, it is always possible to choose an inertial frame of reference in which locally the gravitational field vanishes.  Accordingly, any EMT constructed from the metric and its first derivatives would vanish in such a local frame.  However, if a tensor vanishes in one frame it has to vanish in any other frame, hence hampering the possibility of defining a non-vanishing energy-momentum.  A possible way out is represented by \emph{pseudotensorial} definitions of energy-momentum, such as the Einstein or the Landau-Lifshitz pseudotensors.  The latter, in particular, is symmetric (\ie\ angular momentum is conserved) and vanishes in any locally inertial frame; moreover, its sum with the matter energy-momentum has a vanishing divergence.

On a mathematical level, the impossibility to define an energy-momentum for the gravitational field is due to the fact that the metric $g_{ab}$ plays the role of background and dynamical field simultaneously.  The variation of the Einstein-Hilbert action with respect to the metric yields the Einstein field equations and there is no other independent variable which would imply the existence of a conserved symmetric tensor \cite{Szabados2009a}.  The next natural choice of variable would be the connection, which is a non-local object since it connects different fibers of the tangent bundle.  As said above, components of the connection can be eliminated locally, but not in an extended region.  This gives hope that one might be able to define a meaningful notion of energy in a quasi-local way.  It turns out that for algebraically special spacetimes there is the possibility of defining a local energy-momentum tensor which measures the energy of the gravitational field in a non-local way, as it depends on second derivatives of the metric.  We are going to analyze the properties of a specific proposal along these lines.

Let us recall at this point that the Riemann tensor can be split into trace (Ricci) and traceless (Weyl) parts.  The former is of course related to the matter content through Einstein's field equations, which establish a pointwise correspondence between the energy of a source distribution and the local behavior of the spacetime; the latter instead, which is the only part that survives in vacuum, describes properties of free gravitational fields and their propagation between distant regions (\eg gravitational waves).  It is hence reasonable to look for intrinsic energetic properties of free gravitational fields in the Weyl part of the Riemann tensor.

In the search for the suitable gravitational EMT it is perhaps useful to enforce a formal analogy that has been pointed out in several occasions between the gravitational field and the electromagnetic one \cite{maartens1998gravito,mashhoon2001gravitoelectromagnetism}, analogy that is based on the correspondence between the Maxwell tensor $F_{ab}$ and the Weyl tensor $C_{abcd}$.  In a covariant $1+3$ splitting of the spacetime the Weyl tensor can be decomposed into \emph{electric} part, $E_{ab}$, and \emph{magnetic} part, $H_{ab}$, which acquire the role of fundamental dynamical quantities alongside with the basic properties of matter (energy density, pressure, etc.).  In fact, in such splitting the Bianchi identities assume a transparently Maxwellian form and it is possible to recognize the Bel-Robinson (BR) tensor as a
\emph{super-energy tensor} for free gravity.  BR is completely symmetric, traceless and conserved in vacuum; in analogy with the electromagnetic counterpart, the symmetries of such object allow to define a super-energy density and a super-Poynting vector.  The physical interpretation of super-energies is still a matter of debate and research \cite{senovilla2000super,garecki2001some,gomez2014conservation}.  

The first problem one encounters in associating, for instance, the completely timelike component of BR to a notion of energy is that the latter has dimensions of energy squared.  In order to soften such interpretative issue, the \emph{square-root of the BR tensor} (SQBR) has been proposed as a possible definition of gravitational EMT \cite{bonilla1997some}.  Apart from having the right energy dimension, the SQBR possesses interesting properties which have been already pointed out in the study of thermodynamic behavior of classical spacetimes, particularly in connection with the notion of \emph{gravitational entropy} \cite{Clifton-2013}.  The definition of entropy arising from this framework enjoys some of the desired properties that one expects from the entropy of the gravitational field \cite{sussman2014gravitational,Acquaviva-2015}: it is non-negative, it vanishes if and only if the Weyl tensor is zero, increases as structures (inhomogeneities) form in the Universere and reduces to the Bekenstein-Hawking entropy for Schwarzschild black holes.  In the present paper we are going to review properties of the SQBR and provide new insights with the aim of clarifying better its possible relation with the thermodynamic behavior of the gravitational field.

In section \ref{sec2:BR} we present the derivation of SQBR in both spinor and tensorial forms for spacetimes that belong to Types N and D in the Petrov classification; we further propose a way to fix the inherent freedom in the definition of the SQBR.  In section \ref{sec3:optical} we present the relations between optical scalars of timelike congruences and spin coefficients, which will be useful in the subsequent analysis.  In section \ref{sec:fluid-interpretation} we exploit the general decomposition of any EMT in order to relate the geometric properties of the congruences to usual thermodynamic quantities; we note that the SQBR is in general not divergence-free even in vacuum, signaling an intrisically dissipative behavior of some gravitational configurations; we further interpret the timelike projection of its covariant divergence as a first law of thermodynamics, expressing the variation of gravitational energy as due both to \emph{work} made on/by the system, and to \emph{dissipation}.  In section \ref{sec5:electro} we show that the SQBR can be recast in the form of an electromagnetic EMT: the components of the SQBR in this setting exactly reproduce those of its EM counterpart.  We apply such schemes in sections \ref{sec:thermo-N} and \ref{sec:thermo-D} to the cases of Type N and Type D spacetimes respectively; we provide specific examples in which the behavior of the thermodynamic quantities is analyzed, highlighting as well the intrinsic observer-dependence of some effects.  Eventually, in section \ref{sec5:concl} we provide a conclusive overview of the analysis and present possible future paths to be undertaken.  Throughout the paper, the metric signature is $(+,-,-,-)$. Tensorial indices will be denoted by $a,b,c,\dots$ while spinor indices will be denoted by $A,B,C, \dots$. The tetrad components
of tensors will be labeled by $\bm{a}, \bm{b}, \bm{c}, \dots$  and the spatial components of tensors by $\bm{i}, \bm{j}, \bm{k}, \dots$. The Riemann tensor is
defined by $2\,\nabla_{[c}\nabla_{d]} X_a = - R_{abcd}X^b$.

\section{The Bel-Robinson tensor and its square root}\label{sec2:BR}

It is interesting to note how most of the properties of the BR tensor mirror those of the Maxwell tensor of electromagnetism.  The formal analogy that ensues has been investigated by several authors and it establishes the role of the BR tensor as a super-energy-momentum for the gravitational field.  While we refer the reader to \cite{maartens1998gravito} for a transparent analysis of the electromagnetic-like properties of the BR tensor, we present here only the notions relevant for our discussion.  For a generic spacetime, the BR tensor is defined as
\begin{equation}
  T_{abcd} = C_{aecf}\, \Cdudu{b}{e}{d}{f} + {}^*\nspace C_{aecf}\, {}^*\nspace \Cdudu{b}{e}{d}{f}\, ,
\end{equation}
and it enjoys the properties of complete symmtery, tracelessness and covariant conservation in vacuum.  In the spirit of the aforementioned analogy, given a generic timelike congruence $u^a$ one can define electric $E_{ac}=C_{abcd}u^bu^d$ and magnetic $H_{ac}=-\, ^*C_{abcd}u^bu^d$ parts of the Weyl tensor.  In terms of these one can write down, for instance, a super-energy density as the completely timelike component of BR:
\begin{equation}
  U \equiv T_{abcd}u^au^bu^cu^d = E_{ef}E^{ef} + H_{ef}H^{ef}\label{superen}\, ,
\end{equation}
which is an invariant quantity under spatial duality rotations.  On one hand, the possibility of defining the energy of the gravitational field through \eqref{superen}, although tempting, is however hampered by the simple fact that a super-energy has dimensions of energy squared.  On the other hand, in \cite{bonilla1997some} the authors notice that the Bel tensor in the case of Einstein-Maxwell systems can be decomposed irreducibly in two parts, one of which is the \emph{square of the electromagnetic EMT}; the second part is the BR tensor.

The two considerations above open the possibility of considering the ``square root'' of Bel-Robinson (SQBR) as a possible definition of EMT for free gravitational fields.  Such proposal has been picked up by \cite{Clifton-2013} and subsequently applied to particular cases in \cite{sussman2014gravitational,Acquaviva-2015}. In what follows we will employ the Newman-Penrose (NP) formalism in which the properties of principal null directions are encoded in the spin coefficients.

We will also employ the formalism of 2-spinors \cite{Penrose-Rindler-I,Stewart-1993} because in this
formalism the algebraic properties of curvature tensors become more transparent. At each point of the spacetime, the space of spinors $S^A$ is a complex two dimensional space whose elements will be labeled by capital letters, \eg $\xi^A$. With $S^A$ we canonically associate the dual space $S_A$, the complex conjugate space $S^{A'}$
and the complex conjugate dual space $S_{A'}$. Recall that any spinor $\xi^A$ gives rise to
a real null spacetime vector,
\begin{eqnarray}
  \label{eq:soldering}
  k^a = \xi^A\,\bar{\xi}^{A'},
\end{eqnarray}
where bar denotes complex conjugation. The space of spinors
is equipped with the symplectic form $\epsilon_{AB}$ which is related to the spacetime metric by
\begin{eqnarray}
  \label{eq:epsilon-metric}
  g_{ab} = \epsilon_{AB}\,\epsilon_{A'B'}.
\end{eqnarray}
The spinorial equivalent of the Weyl tensor $C_{abcd}$ is represented by a totally symmetric spinor $\Psi_{ABCD}$, so that the anti-self-dual part of the Weyl tensor is
\begin{eqnarray}
  \label{eq:asd-weyl}
  \CC_{abcd} = \Psi_{ABCD}\,\epsilon_{A'B'}\,\epsilon_{C'D'}
\end{eqnarray}
and the real Weyl tensor is
\begin{eqnarray}
  \label{eq:weyl-spinor}
  C_{abcd} = \CC_{abcd} + \bar{\CC}_{abcd}.
\end{eqnarray}
Similarly, the anti-self-dual part of the electromagnetic field $F_{ab} = \FF_{ab} + \bar{\FF}_{ab}$ is given by a symmetric spinor $\phi_{AB}$ as
\begin{eqnarray}
  \label{eq:asd-maxwell}
  \FF_{ab} = \phi_{AB}\,\epsilon_{A'B'}.
\end{eqnarray}

Let us briefly review the spinorial construction of SQBR, following \cite{Bergqvist-1998}, and discuss several related issues. As any totally
symmetric spinor, $\Psi_{ABCD}$ can be factorized into the symmetrized product of four univalent spinors called \emph{principal spinors}, 
\begin{eqnarray}
  \label{eq:weyl-factorization}
  \Psi_{ABCD} = \alpha_{(A}\beta_B\gamma_C\delta_{D)}.
\end{eqnarray}
Each principal spinor determines a \emph{principal null direction}, \eg principal spinor $\alpha_A$ gives rise to a principal null vector $\alpha^A\,\bar{\alpha}^{A'}$ and corresponding null direction.
The Petrov Type of the Weyl tensor is then determined by the number of linearly independent principal null directions. 

The spinor equivalent of the Bel-Robinson tensor is
\begin{eqnarray}
  \label{eq:BR-spinors}
  T_{abcd} = \Psi_{ABCD}\,\bar{\Psi}_{A'B'C'D'}
\end{eqnarray}
and is manifestly trace-free. Moreover, by the spinor form of Bianchi identities in vacuum,
\begin{eqnarray}
  \label{eq:Bianchi-spinor}
  \nabla_{A'}^A\Psi_{ABCD}=0,
\end{eqnarray}
the tensor $T_{abcd}$ is also divergence-free in vacuum. A symmetric tensor $t_{ab}$
is called \emph{square root of the Bel-Robinson tensor} (SQBR), if it satisfies
\begin{eqnarray}
  \label{eq:SQBR-def}
  T_{abcd} = t_{(AB(A'B'}\,t_{C'D')CD)}.
\end{eqnarray}
Clearly, if $t_{ab}$ is a SQBR, then also
\begin{eqnarray}
  \label{eq:SQBR-freedom}
  t_{ab} + f\,g_{ab} \equiv t_{ABA'B'} + f\,\epsilon_{AB}\,\epsilon_{A'B'}
\end{eqnarray}
is a SQBR and we will discuss appropriate choice of $f$ in Section \ref{sec:constraints}. The question is whether, apart from freedom given by (\ref{eq:SQBR-freedom}), one can unambiguously
define $t_{ab}$ in terms of principal spinors of the Weyl spinor. It turns
out that this is feasible only in spacetimes of Types N and D.

\subsection{Type N spacetimes}
\label{sec:SQBR-N}

In Type N spacetimes, the Weyl tensor admits a single degenerate principal null direction, that is
\begin{eqnarray}
  \label{eq:weyl-type-N}
  \Psi_{ABCD} = \alpha_{A}\alpha_B\alpha_C\alpha_{D}.
\end{eqnarray}
Defining
\begin{eqnarray}
  \label{eq:N-phiAB}
  \phi_{AB} = \alpha_A\,\alpha_B,
\end{eqnarray}
the SQBR satisfying (\ref{eq:SQBR-def}) can be easily constructed as
\begin{eqnarray}
  \label{eq:N-SQBR}
  t_{ab} = \phi_{AB}\,\bar{\phi}_{A'B'}.
\end{eqnarray}
In order to find a convenient tensorial representation we first introduce an appropriate spin basis. The vacuum Bianchi identities (\ref{eq:Bianchi-spinor}) applied to
the Weyl spinor of the form (\ref{eq:weyl-type-N}) imply
\begin{eqnarray}
  \label{eq:N-Bianchi}
  \alpha^A \nabla_{AA'} \alpha_B = \pi_{A'}\,\alpha_B,
\end{eqnarray}
for some spinor $\pi_{A'}$, showing that the null direction given by $\alpha^A$ is a (non-affinely parametrized) geodesic with vanishing shear. Let us define the spinor
\begin{eqnarray}
  \label{eq:N-oA}
  o^A = \chi\,\alpha^A, \qquad \chi = \exp\left\{ - \int \bar{\alpha}^{A'}\,\pi_{A'}\,\dd v \right\},
\end{eqnarray}
where the integral is taken along the orbit of $\alpha^A \bar{\alpha}^{A'}$, which is parallelly propagated along $\ell^a = o^A\bar{o}^{A'}$, \ie
\begin{eqnarray}
  \label{eq:N-oA-prop}
  D o^A = 0\,.
\end{eqnarray}
We complete $o^A$ to the spin basis by introducing a spinor $\iota^A$ normalized by $o_A \iota^A = 1$ and propagate it by the condition $D\iota^A = 0$. In this basis, the Weyl spinor (\ref{eq:weyl-type-N})
reads
\begin{eqnarray}
  \label{eq:N-weyl-basis}
  \Psi_{ABCD} = \Psi_4\,o_A\,o_B\,o_C\,o_D, \qquad \Psi_4 = \chi^{-4},
\end{eqnarray}
and the following spin coefficients vanish:
\begin{eqnarray}
  \label{eq:N-spin-coeffs-1}
  \kappa = \eps = \pi = \sigma = 0,
\end{eqnarray}
where the shear $\sigma$ vanishes by virtue of the Bianchi identities. The remaining non-trivial Bianchi identities are
\begin{eqnarray}
  \label{eq:bianchi-N}
  D\Psi_4 = \rho\,\Psi_4, \qquad
  \delta\Psi_4 = (\tau - 4\beta)\Psi_4. 
\end{eqnarray}
The null tetrad $(\ell^a, n^a, m^a, \bar{m}^a)$ induced by the spin basis introduced above is covariantly constant along the principal null direction of the Weyl tensor, \ie
\begin{eqnarray}
  \label{eq:N-D-tetrad}
  D \ell^a = D n^a = D m^a = 0.
\end{eqnarray}
Now we can write
\begin{eqnarray}
  \label{eq:N-phiAB-basis}
  \phi_{AB} = \phi_2\,o_A\,o_B, \qquad \phi_2 = \sqrt{\Psi_4},
\end{eqnarray}
and the SQBR (\ref{eq:N-SQBR}) reads
\begin{eqnarray}
  \label{eq:N-SQBR-basis}
  t_{ab} =\phi_{AB}\,\bar{\phi}_{A'B'} =  |\Psi_4| \ell_a\,\ell_b.
\end{eqnarray}
Taking the freedom (\ref{eq:SQBR-freedom}) into account, we conclude that the most general form of SQBR in Type N spacetimes reads
\begin{eqnarray}
  t_{ab} &= c\,|\Psi_4| \ell_a\,\ell_b + f\,g_{ab},
\end{eqnarray}
where $\ell^a$ is the affinely parametrized principal null direction of the spacetime and $f$ is an arbitrary function; the numerical dimensionless constant $c$ has been put in by hand in order
to make comparison with different possible conventions. 

\subsection{Type D spacetimes}
\label{sec:SQBR-D}

Type D spacetimes admit two linearly independent principal null directions so that the Weyl spinor acquires the form
\begin{eqnarray}
  \label{eq:weyl-D}
  \Psi_{ABCD} = \alpha_{(A}\beta_B\alpha_C\beta_{D)}.
\end{eqnarray}
Following the same procedure as in Type N we introduce spinor basis such that $o^A$ and $\iota^A$ are proportional to $\alpha^A$ and $\beta^B$, respectively, and associated $\ell^a$ and $n^a$
are shear-free null geodesics ($\kappa=\nu=\lambda=\sigma=0$). In this case, however, we cannot affinely parametrize both and hence we keep the general parametrization so that the spin coefficients
$\eps$ and $\gamma$ are not restricted.

There is a unique way of factorizing $\Psi_{ABCD}$ in the form
\begin{eqnarray}
  \label{eq:D-factor}
  \Psi_{ABCD} = \phi_{(AB}\phi_{CD)},
\end{eqnarray}
where
\begin{eqnarray}
  \label{eq:D-phiAB-basis}
  \phi_{AB} = -2\,\phi_1\,o_{(A}\,\iota_{B)} \qquad \mathrm{and}~\phi_1 = \sqrt{\Psi_2}.
\end{eqnarray}
Now, the most general form of $t_{ab}$ turns out to be
\begin{eqnarray}
  t_{ab} &= 2\,c\,|\Psi_2| \left( \ell_{(a}\,n_{b)} + m_{(a}\,\bar{m}_{b)}\right) + f\,g_{ab},
\end{eqnarray}
where $m^a$ and $\bar{m}^a$ complete the null tetrad and again $f$ is an arbitrary function. 

As a final comment regarding the construction of the SQBR, notice that for Weyl spinors \eqref{eq:weyl-factorization} of Types different from N and D, there is no unambiguous way to pair up the principal spinors in order to form the spinor $\phi_{AB}$.  Hence the definition of EMT in those cases would not be unique.

\subsection{Constraints on the SQBR}
\label{sec:constraints}

Notice that although vectors $\ell^a$ and $n^a$ are unique only up to rescaling $\ell^a \rightarrow A \,\ell^a$, $n^a \rightarrow A^{-1} n^a$ with arbitrary function $A$, the SQBR in both Types is, in fact,
invariant under such transformation. Hence, the only true freedom in the definition of SQBR lies in the choice of the arbitrary function $f$ which has to be specified
and constrained.

In previous works \cite{Clifton-2013,Acquaviva-2015}, authors have enforced energy-momentum
conservation to do so, hence finding the form of $f$ such that the timelike projection
of the covariant divergence of the EMT vanishes, \ie $u_a \nabla_b t^{ab}=0$. We have seen that
Type D and N spacetimes have different definitions of EMT: consequently, the form of
$f$ consistent with the above condition is different for the two Types. We argue that,
although energy conservation can be considered a natural requirement, such condition is
not entirely justified and that a different constraint, motivated by different considerations,
could be equally enforced.

In the theory of General Relativity, gravity propagates at the speed of light. This
theoretical assertion has been tested in several circumstances \footnote{For a review on experimental confirmations of GR see \cite{will2014confrontation}.}, for instance through the measurement of the deflection of light inside the Solar System \cite{fomalont2003measurement} and in the recent
observation of gravitational wave emission in concomitance with gamma ray burst
emission from a neutron star merger \cite{abbott2017gravitational}.
This implies that, within a small error margin, the hypotetical particle carrier of the gravitational field should
be massless. The energetic properties of any field expressed by a massless particle are
described by a traceless EMT. Consequently, in order to convey the aforementioned facts,
we conjecture that the gravitational EMT should be traceless. Formally, the function $f$
measures the deviation of the EMT from tracelessness, so our conjecture amounts to
choosing $f = 0$. Such condition is applicable in both Type D and N simultaneously, in
agreement with the reasonable expectation that the speed of propagation of gravitational
effects should be independent from the Type of spacetime.

At this point one could say: we have now an EMT consistent with the masslessness
of a would-be graviton; what about energy conservation? Is it not equally important? For
a generic fluid in presence of gravity, one can rest assured that the covariant divergence
of its EMT vanishes, thanks to Einstein's field equations and Bianchi identities. An EMT
for the gravitational field cannot rely on such promise, as it is not supposed to enter in the
right-hand side of Einstein's equations. Hence one has to deal with the possibility of energy dissipation
and its interpretation in this context. Such effects are here expressed in terms of properties
of the null congruences, hence they primarily depend on the symmetries of the spacetime.
We argue further that the presence or otherwise of such dissipative effects depends on
the choice of the congruence of observers. Such feature is generally present in standard
fluids, where an observer might experience dissipative effects just because it is not
comoving \cite{coley1983zero,herrera2011tilted}. This implies that in some spacetimes one could identify “generalized Killing
observers” for which the gravitational energy is conserved. Ultimately, such energy flows
can be ascribed to the state of the observer with respect to the geometric properties of
the principal null congruences.

Notice that for matter fields tracelessness of the EMT implies conformal invariance of the equations of motion: this is not the case for SQBR, since any conformal transformation of the vacuum field equations generates a Ricci curvature,  so that the conformally rescaled spacetime does not satisfy the vacuum equations anymore.  Thus, our choice of a traceless EMT is not motivated by a requirement of conformal invariance.

\section{Optical scalars for timelike congruences}
\label{sec3:optical}

In this paper we consider only spacetimes of Type N and Type D for which the SQBR can be defined unambiguously. Up to the freedom to be discussed, the algebraic structure
of these spacetimes gives rise to a preferred family of timelike congruences (parametrized by the boost in the plane defined by $\ell^a$ and $n^a$) with the unit
tangent vector field $u^a$. In what follows we will refer
to the decomposition of the tangent bundle $TM$ into the subspace $T^uM$ parallel
to $u^a$ and the orthogonal space $T^\perp M$ as the \emph{1+3 decomposition}, \ie we write $TM = T^u M \oplus T^\perp M$, where the projector $h : TM \mapsto T^\perp M$ is defined by
\begin{eqnarray}
  h^a_b &= \delta^a_b - u^a\,u_b.
\end{eqnarray}
Notice that since $u^a$ is not, in general, hypersurface-orthogonal, the decomposition on the level of tangent spaces does not yield the foliation of the spacetime, \ie\ the 3+1 decomposition.

Let us decompose the covariant derivative of $u^a$ in a standard way as
\begin{eqnarray}
  \label{eq:ua-decomp}
  \nabla_a u_b &= u_a\,\dot{u}_b + \sigma_{ab} + \frac{1}{3}\,h_{ab}\,\theta + \omega_{ab},
\end{eqnarray}
where $\dot{} \equiv u^c\nabla_c$ and the expansion $\theta$, shear $\sigma_{ab}$ and twist $\omega_{ab}$ are defined by
\begin{eqnarray}
  \label{eq:optical-tensors}
  \theta = h^{ab} \nabla_a u_b, \qquad
  \sigma_{ab} = h_a^c\,h_b^d \nabla_{(c}u_{d)} - \frac{1}{3}\,h_{ab}\,\theta, \qquad
  \omega_{ab} = h^c_a\,h^d_b \nabla_{[c}u_{d]}.
\end{eqnarray}
For later convenience we also define the spatial part of the spacetime connection by the usual relation
\begin{eqnarray}
  \label{eq:DD-def}
\DD_a X_b = h_a^c h_b^d \nabla_c X_d
\end{eqnarray}
for any $X_a$. 
A natural choice of unit timelike vector adapted to the principal null directions $\ell^a$ and $n^a$ is given by
\numparts
\begin{eqnarray}
  \label{eq:ua-boost}
  u^a = \frac{1}{\sqrt{2}}\left(A \, \ell^a + A^{-1}\,n^a\right),
\end{eqnarray}
where the boost parameter $A$ is an arbitrary function of spacetime coordinates.  We will see later on that, in terms of such preferred family of observers, the thermodynamic quantities associated to the gravitational field acquire the  simplest form.  For instance, in the fluid interpretation of $t_{ab}$ to be given later, in Type D spacetimes this congruence corresponds to observers comoving with a fluid.

We can complement $u^a$ adapted to the principal null directions with the triad
\begin{eqnarray}
 x^a& = \frac{1}{\sqrt{2}}\left( m^a + \bar{m}^a \right)\, , \quad 
 y^a = \frac{1}{\sqrt{2}\, \ii}\left( m^a - \bar{m}^a \right)\, , \\
 z^a &= \frac{1}{\sqrt{2}}\left(  A\,\ell^a - A^{-1}\,n^a \right)\, , \label{eq:tetradz}
\end{eqnarray}
\endnumparts
such that $\{ u^a, x^a, y^a, z^a \}$ form an orthonormal tetrad. Occasionally we will employ the components of tensors with respect to an orthonormal tetrad. For that reason we introduce the
soldering form $e^a_{\bm{a}}$ where $a$ is an abstract index and $\bm{a}=0,1,2,3$ is a concrete index labeling the elements of the tetrad, \ie
\begin{eqnarray}
  \label{eq:soldering-form}
  e^a_{\bm{a}}  = ( u^a, x^a, y^a, z^a ).
\end{eqnarray}
Tetrad components of tensor $X^{a\dots b}_{c\dots b}$ will be denoted by
\begin{eqnarray}
  \label{eq:tetrad-components}
  X^{\bm{a}\dots \bm{b}}_{\bm{c}\dots \bm{d}} = e^{\bm{a}}_a \cdots e^{\bm{b}}_b\,e_{\bm{c}}^c \cdots e_{\bm{d}}^d\,X^{a \dots b}_{c \dots b},
\end{eqnarray}
where $e^{\bm{a}}_a = (u_a, - x_a, -y_a, -z_a)$ is dual to $e^a_{\bm{a}}$. For labeling the elements of the spatial triad we will employ indices $\bm{i}=1,2,3$.

In terms of the spin coefficients, the expansion of the congruence of $u^a$ has the form
\begin{eqnarray}
  \label{eq:expansion}
  \theta &=  \frac{1}{\sqrt{2}A}\left( A^2 \left(\eps+\bar{\eps}\right) - \gamma - \bar{\gamma} - A^2 \left(\rho + \bar{\rho}\right) + \mu + \bar{\mu} +A\, D A-\frac{\Delta A}{A}\right),
\end{eqnarray}
the non-vanishing projections of the shear tensor $\sigma_{ab}$ onto the orthonormal tetrad (\ref{eq:ua-boost}--\ref{eq:tetradz}) read
\numparts
\begin{eqnarray}
  \label{eq:shear-comps}
  \sigma_{\bm{11}} &= \frac{1}{3}\,\theta +\frac{1}{2\sqrt{2}}\left[ A\left( \rho+\bar{\rho}+\sigma+\bar{\sigma} \right)-A^{-1}\left( \mu+\bar{\mu}+\lambda+\bar{\lambda} \right) \right] , \\
  \sigma_{\bm{12}} &= -\frac{\ii}{2\sqrt{2}\,A}\, \left[ \lambda-\bar{\lambda}+A^2 \left( \sigma-\bar{\sigma} \right) \right], \\
  \sigma_{\bm{13}} &= - \frac{1}{4\sqrt{2}}\left[ 2\left( \alpha+\bar{\alpha}+\beta+\bar{\beta} \right) +\pi+\bar{\pi}+\tau+\bar{\tau} -A^2 (\kappa+\bar{\kappa})  \right.\\
  &\qquad\qquad \left.-A^{-2}(\nu+\bar{\nu}) +2A^{-1}(\delta A+\bar{\delta}A) \right], \\
  \sigma_{\bm{23}} &= - \frac{\ii}{4\sqrt{2}}\left[ 2\left( \alpha-\bar{\alpha}-\beta+\bar{\beta} \right) + \pi-\bar{\pi}-\tau+\bar{\tau} +A^2(\kappa-\bar{\kappa}) \right.\\
  &\qquad\qquad\left. -A^{-2}(\nu-\bar{\nu}) -2A^{-1}(\delta A- \bar{\delta}A) \right], \\
  \sigma_{\bm{33}} &= - \sigma_{\bm{11}} - \sigma_{\bm{22}} = - 2\,\sigma_{\bm{11}},
\end{eqnarray}
\endnumparts
and orthonormal components of the twist are
\numparts
\begin{eqnarray}
  \label{eq:twist-comps}
  \omega_{\bm{12}} &= - \frac{\ii}{2\sqrt{2}}\left( A(\rho-\bar{\rho}) +A^{-1} (\mu - \bar{\mu}) \right), \\
\omega_{\bm{13}} &= -\frac{1}{4\sqrt{2}}\left[ 2 \left( \alpha+\bar{\alpha}+\beta+\bar{\beta} \right) -\pi-\bar{\pi}-\tau-\bar{\tau} +A^2(\kappa+\bar{\kappa}) \right.\\
&\qquad\qquad \left.+A^{-2}(\nu+\bar{\nu})+2A^{-1}(\delta A+\bar{\delta}A) \right],\\
\omega_{\bm{23}} &= -\frac{\ii}{4\sqrt{2}}\left[ 2(\alpha-\bar{\alpha}-\beta+\bar{\beta}) -\pi+\bar{\pi}+\tau-\bar{\tau}-A^2(\kappa-\bar{\kappa}) \right.\\
&\qquad\quad \left. +A^{-2} (\nu-\bar{\nu}) -2A^{-1} (\delta A-\bar{\delta} A) \right]\, .
\end{eqnarray}
Finally, the components of the acceleration are given by
\begin{eqnarray}
  \label{eq:udot-comps}
  \dot{u}_{\bm{1}} &= \frac{1}{2\sqrt{2}}\left( A^2(\kappa+\bar{\kappa}) - A^{-2}(\nu+\bar{\nu})+\tau+\bar{\tau}-\pi-\bar{\pi} \right)\,, \\
  \dot{u}_{\bm{2}} &= \frac{\ii}{2\sqrt{2}}\left( A^2(\bar{\kappa}-\kappa) + A^{-2}(\bar{\nu}-\nu)-\tau+\bar{\tau}-\pi+\bar{\pi} \right)\,,\\
  \dot{u}_{\bm{3}} &= - \frac{1}{\sqrt{2}}\left( DA+A^{-2}\,\Delta A + A^{-1}(\gamma+\bar{\gamma})+A(\eps+\bar{\eps})\right).
\end{eqnarray}

\endnumparts
Since the components of the twist are, in general, non-vanishing, the congruence $u^a$ is not hypersurface orthogonal and therefore it does not define a foliation of the spacetime corresponding to a family of observers.  Can we choose $A$ so as to make $u^a$ hypersurface orthogonal? Non-trivial projections of
the condition $u_{[a}\nabla_bu_{c]}=0$ read
\begin{eqnarray}
  \label{eq:hypersurface-ortho-conds}
  \delta \log A = \frac{1}{2}(\tau+\bar{\pi})-(\bar{\alpha}+\beta),
  \qquad
  A^2 =\frac{\mu - \bar{\mu}}{\bar{\rho}-\rho}.
\end{eqnarray}
Interestingly, one can check that integrability conditions for system \eqref{eq:hypersurface-ortho-conds} are formally satisfied. However, the function $A$ is imaginary unless the fraction is positive.
This happens, for example, below the Kerr horizon but \emph{not} outside the horizon. 

\section{Fluid-like description of the EMT for generic observers}
\label{sec:fluid-interpretation}

At this point, we are still missing the interpretation and the dynamical behavior of the SQBR.  Such interpretation is possible only after we specify the family of observers, whose four-velocity defines a natural splitting of all tensorial quantities into spatial and temporal parts, which brings us naturally to a fluid-like description.

Consider a general timelike congruence $t^a$, with $t^a t_a = 1$, which will be interpreted as a set of worldlines representing a family of observers with four-velocity $t^a$. With $t^a$ we associate the orthogonal
projector $h_{ab} = g_{ab}- t_a\,t_b$ and any tensor can be decomposed into parts parallel and orthogonal to $t^a$, respectively. In particular, any energy-momentum tensor $t_{ab}$ can
be decomposed as
\begin{eqnarray}
  \label{eq:tab-decomp}
  t_{ab} = \muG\,t_a\,t_b +2\,q_{(a}\,t_{b)} + \PG\,h_{ab} + \pi_{ab},
\end{eqnarray}
where $\muG$ is interpreted as the energy density measured by observer $t^a$, $q_a$ is a purely spatial vector representing the heat flux, $\PG$ is the isotropic pressure
and $\pi_{ab}$ is a purely spatial, trace-free symmetric tensor representing anisotropic pressure.  Notice that these quantities behave like parts of a tensor under Lorentz transformations of $t_a$ (see \ref{app:lorentz}), rather than like scalars, vectors or tensors.  This is true in this context as well as in electromagnetism, and it means that some components of the EMT can be eliminated by choosing an appropriate frame of reference.
%

For normal matter the EMT is covariantly conserved; however, for the gravitational EMT we have
\begin{eqnarray}
  \label{eq:Fa-def}
  \nabla_b t^{ab} = -F^a &\, .
\end{eqnarray}
In analogy with the electromagnetic case, where $\nabla_a T^{ab}=- F^{bc}j_c$, $F_a$ can be interpreted as a force density.  

We will show now that the timelike component of \eqref{eq:Fa-def} can be recast in the form of a first law of thermodynamics, namely
\begin{eqnarray}
  \label{eq:xia-Ja}
  t^b \nabla^a t_{ab} = \Grav{\dot{\mu}} + (\muG-\PG)\theta_t + \nabla^a q_a - q_b\,\dot{t}^b - \pi_{ab}\,\sigma^{ab}\, .
\end{eqnarray}
Let us denote by $\bm{\epsilon}$ the four-dimensional volume form $\epsilon_{abcd}$, so that the three-dimensional volume form for
observers with four-velocity $t^a$ is
\begin{eqnarray}
  \label{eq:3d-volume}
  \bm{\omega}_t = i_t \bm{\epsilon}, \qquad \mathrm{or} \qquad (\omega_t)_{bcd} = t^a \,\epsilon_{abcd}.
\end{eqnarray}
One would like to analyze the most generic variations of the thermodynamical quantities which would include perturbations of the metric as well as instantaneous deformations of the volume analogous to virtual displacements. In the present work we restrict to variations along the chosen timelike congruence $t^a$, so that the variations will correspond to Lie derivatives $\Lie_t$. For instance,
the three-volume will vary as
\begin{eqnarray}
  \label{eq:Lie-omega}
  \Lie_t \bm{\omega}_t = \theta\,\bm{\omega}_t \,,
\end{eqnarray}
where $\theta = \nabla_a t^a$ is the expansion of $t^a$. This term would correspond to $\delta V$ in a first law, but it does not measure the deformations
of the shape of the volume which are related to shear of the congruence. Both these effects are encoded in the Lie derivative of $h_{ a b}$,
\begin{eqnarray}
  \label{eq:lie-hab}
  \frac{1}{2}\Lie_t h_{ ab } = \sigma_{ ab} + \frac{1}{3}\,h_{ ab }\,\theta_t \equiv \delta V_{ab}
\end{eqnarray}
which is a purely spatial tensor. At the same time, we define the variation of the energy contained in an infinitesimal volume as
\begin{eqnarray}
  \label{eq:energy-variation}
  \Lie_{t} (\muG\,\bm{\omega}_t) = (\dot{\mu}_G + \theta_t\,\muG) \bm{\omega}_t \equiv \delta U\,\bm{\omega}_t.
\end{eqnarray}
Combining (\ref{eq:lie-hab}) and (\ref{eq:energy-variation}) we can rewrite (\ref{eq:xia-Ja}) in the following form:
\begin{eqnarray}
  \label{eq:1st-law}
  \delta U =  \delta Q - \delta W - F_a\,t^a,
\end{eqnarray}
where we have denoted
\begin{eqnarray}
  \label{eq:delta-Q-W}
  \delta W = - t^{ab} \delta V_{ab},
  \qquad
  \delta Q = - \DD_a q^a + 2\,\dot{t}^a\,q_a. 
\end{eqnarray}
Clearly, $\delta W$ is a generalized work term which includes not only the change of the volume but also its deformation due to shear. The term $\delta Q$ originates from the
presence of the heat flux $q^a$ in a given frame and hence we interpret it as the dissipative part of the first law. The presence of this term is in general observer
dependent and, in fact, we will show that $\delta Q=0$ for particular observers in general Type D spacetimes. Finally, the last term in (\ref{eq:1st-law}) amounts to
an intrinsic energy dissipation and we interpret its presence as the indication that even the free gravitational field is not a truly isolated system. Nevertheless, since $F^a$ is
a spatial vector, its temporal projection $F_a t^a$ can be set to zero by an appropriate choice of the observer; in such case, the intrinsic dissipation will contribute to the spatial projections
of the balance equation (\ref{eq:Fa-def}) only.  The presence of intrinsic dissipation in the gravitational sector of GR as well as in some modified gravity theories has been
pointed out, \eg, by \cite{chirco2010nonequilibrium}: the authors, in particular, interpret such dissipation as arising from work done upon the microscopic degrees of freedom of gravity.
We do not enter here in the details of such interpretation, although it is clear that the possibility of describing thermodynamically the macroscopic behavior of a system can hint towards
a corresponding microscopic level of description and its statistical mechanical features.

In sections \ref{sec:thermo-N} and \ref{sec:thermo-D} we will provide explicit form of the terms in \eqref{eq:1st-law} in particular cases of Type N and Type D spacetimes and discuss their behavior in different frames.  Before that, in the next section we are going to present an electromagnetic formulation of free gravity.

\section{Electromagnetic interpretation}
\label{sec5:electro}

Analogies between gravitational and electromagnetic fields are well-known and they are systematically exploited
in the formalism of gravito-electromagnetism \cite{maartens1998gravito}. Here we pursue the observation
that if the SQBR can be defined, it naturally gives rise to an electromagnetic field with anti-self-dual part (\ref{eq:asd-maxwell})
where the spinor $\phi_{AB}$ is now given either by (\ref{eq:N-phiAB-basis}) for Type N or by (\ref{eq:D-phiAB-basis}) for Type D spacetime. In both cases,
apart from freedom (\ref{eq:SQBR-freedom}), the SQBR has the form
\begin{eqnarray}
  \label{eq:tab-gen-EM}
  t_{ab} = \phi_{AB}\,\bar{\phi}_{A'B'} = \FF_{ac}\,\Tud{\bar{\FF}}{c}{b},
\end{eqnarray}
which exactly corresponds to the energy-momentum tensor of electromagnetic field in otherwise empty spacetime \cite{Penrose-Rindler-I}. Using the following identities
\begin{eqnarray*}
  6\,\phi_{(AB}\,\phi_{CD)} = 4\,\phi_{AB}\,\phi_{CD} + 2\,\phi_{A(C}\phi_{D)B} + \Phi^2\,\epsilon_{A(C}\,\epsilon_{D)B}, \\
  2\,\FF_{a[b}\,\FF_{c]d} = \frac{1}{2}\,\epsilon_{AD}\,\epsilon_{BC}\,\epsilon_{A'(B'}\,\epsilon_{C')D'} \Phi^2 - \phi_{A(B}\phi_{C)D}\,\epsilon_{A'D'}\,\epsilon_{B'C'}, \\
  2\,g_{a[b}\,g_{c]d} = - \epsilon_{A(B}\,\epsilon_{C)D} \,\epsilon_{A'D'}\,\epsilon_{B'C'} - \epsilon_{A'(B'}\,\epsilon_{C')D'}\,\epsilon_{AD}\,\epsilon_{BC},
\end{eqnarray*}
we can derive the tensorial form of the anti-self-dual part of the Weyl tensor:
\begin{eqnarray}
  \label{eq:Cabcd-Fab}
  \CC_{abcd} = \phi_{(AB}\phi_{CD)}\epsilon_{A'B'}\,\epsilon_{C'D'} = \frac{2}{3} \left( \FF_{ab}\,\FF_{cd} - \FF_{a[c}\,\FF_{d]b} \right) - \frac{1}{3}\,g_{a[c} g_{d]b} \,\Phi^2,
\end{eqnarray}
where we have denoted
\begin{eqnarray}
  \label{eq:Phi2}
  \Phi^2  = \frac{1}{2}\,\FF_{ab}\,\FF^{ab}.
\end{eqnarray}
We define the electric and magnetic fields as
\begin{eqnarray}
 E_a = F_{ab}t^b,\quad B_a = \Fdual_{ab}t^b\, ,
\end{eqnarray}
in terms of which the anti-self-dual form reads
\begin{eqnarray}
 \FF_{ab} = \frac{1}{2}\, \left( 2\, E_{[a}t_{b]} + \epsilon_{abc}B^c \right) + \frac{\ii}{2}\, \left( 2\, B_{[a}t_{b]} - \epsilon_{abc}E^c \right)\, ,
\end{eqnarray}
where $\epsilon_{abc}\equiv\epsilon_{dabc}t^d$ is the 3-dimensional volume form.  The electric and magnetic parts of the Weyl tensor are then
\begin{eqnarray}
 E_{ab} &= \frac{1}{2}\left( E_aE_b - B_aB_b \right) - \frac{1}{3}\left( E^2-B^2 \right)h_{ab}\, ,\\
 H_{ab} &= E_{(a}B_{b)} -\frac{2}{3} E_cB^c\, h_{ab}\, .
\end{eqnarray}
Now we can relate \eqref{eq:tab-gen-EM} to the irreducible parts of \eqref{eq:tab-decomp} and express the fluid-like quantities in terms of the fields:
\begin{eqnarray}
 \muG &= -\frac{1}{4} \left( E^2 + B^2 \right)\, ,\\
 \PG &= \frac{1}{12} \left( E^2 + B^2 \right)\, ,\\
 q_a &= -\frac{1}{2}\,  \epsilon_{abc} E^bB^c  \equiv - \frac{1}{2}\left( E \times B \right)_a\, ,\\
 \pi_{ab} &= -\frac{1}{2} \left( E_aE_b +B_aB_b \right) +\frac{1}{6} \left( E^2+B^2 \right)h_{ab}\, .
\end{eqnarray}
Notice that these quantities are formally identical to their electromagnetic counterparts: in particular the heat flux $q_a$ is analogous to the Poynting vector.

\section{Thermodynamics of Type N spacetimes}
\label{sec:thermo-N} 

In Type N spacetimes, specializing to the congruence $t^a\equiv u^a$ given by \eqref{eq:ua-boost}, the electromagnetic spinor acquires the form $\phi_{AB}= \phi_2 \,o_A o_B$, cf.\ \eqref{eq:N-phiAB-basis},
and its only non-vanishing component reads
\begin{eqnarray}
 \phi_2 = -E_x-\ii\, B_x = -B_y+\ii\, E_y = \sqrt{\Psi_4}\, .
\end{eqnarray}
We see that in this case the Poynting vector is given by
\begin{eqnarray}
 q_a = - \left( E_x^2+B_x^2 \right) z_a\, .
\end{eqnarray}
Since the electromagnetic field is algebraically special and $E_a B^a = 0$ is an invariant, no frame in which $(E\times B)_a = 0$ can be found and hence the heat flux $q_a$ is always
non-zero.  This feature can be related to the fact that Type N spacetimes describe propagating gravitational radiation, whose presence is not observer-dependent: consequently, such component of the EMT cannot be eliminated by a Lorentz transformation.

The thermodynamic quantities in terms of $\Psi_4$, with respect to frame adapted to the principal null directions, read
\numparts
\begin{eqnarray}
  \label{eq:thermo-N}
  \muG &=& \frac{c}{2}\, A^{-2}\, |\Psi_4|, \\
  \PG &=& - \frac{c}{6}\, A^{-2}\, |\Psi_4|, \\
  q_a &=& \frac{c}{2}\, A^{-2}\, |\Psi_4|\,z_a, \\
  \pi_{ab} &=& - \frac{c}{6}\, A^{-2}\,  |\Psi_4| \left( x_a\,x_b+y_a\,y_b - 2\,z_a\,z_b\right).
\end{eqnarray}
\endnumparts
Under general Lorentz transformation these quantities transform according to the following relations:
\numparts
\begin{eqnarray}
  \label{eq:lorentz-N}
  \Grav{\tilde{\mu}} &=& \muG\,(\Lambda^{\bm{0}}_{\bm{0}})^2 - 2\,q_{\bm{3}}\,(\Lambda^{\bm{0}}_{\bm{0}})(\Lambda^{\bm{3}}_{\bm{0}}) -
                  \PG\left( (\Lambda^{\bm{1}}_{\bm{0}})^2+(\Lambda^{\bm{2}}_{\bm{0}})^2+(\Lambda^{\bm{3}}_{\bm{0}})^2 \right) \nonumber \\
              &&+ \pi_{\bm{11}}\left( (\Lambda^{\bm{1}}_{\bm{0}})^2+(\Lambda^{\bm{2}}_{\bm{0}})^2-2(\Lambda^{\bm{3}}_{\bm{0}})^2\right)\,, \\
  \Grav{\tilde{P}} &=& - \frac{1}{3}\,\Grav{\tilde{\mu}}\,, \label{eosN} \\
  \tilde{q}_{\bm{i}} &=&  \muG\,\Lambda^{\bm{0}}_{\bm{0}}\,\Lambda^{\bm{0}}_{\bm{i}} +2\,q_{\bm{3}}\,\Lambda^{\bm{0}}_{\bm{i}}\,\Lambda^{\bm{3}}_{\bm{0}} -
                         \PG \sum_{{\bm j}=1}^3 \Lambda^{\bm{j}}_{\bm{0}}\,\Lambda^{\bm{j}}_{\bm{i}} \nonumber \\
              && + \pi_{\bm{11}}\left( \Lambda^{\bm{1}}_{\bm{i}} \Lambda^{\bm{1}}_{\bm{0}} +  \Lambda^{\bm{2}}_{\bm{i}} \Lambda^{\bm{2}}_{\bm{0}} -2\,  \Lambda^{\bm{3}}_{\bm{i}} \Lambda^{\bm{3}}_{\bm{0}}\right), \\
  \tilde{\pi}_{\bm{ij}} &=& (\muG-\PG)\Lambda^{\bm{0}}_{\bm{i}}\,\Lambda^{\bm{0}}_{\bm{j}} - (\tilde{P}_{\mathrm{G}}-\PG)\,\delta_{\bm{ij}} + 2\,q_{\bm{3}}\,\Lambda^{\bm{0}}_{(\bm{i}}\,\Lambda^{\bm{3}}_{\bm{j})} \nonumber \\
&&  +\pi_{\bm{11}}\left(  \Lambda^{\bm{1}}_{\bm{i}} \Lambda^{\bm{1}}_{\bm{j}} +  \Lambda^{\bm{2}}_{\bm{i}} \Lambda^{\bm{2}}_{\bm{j}} -2\,  \Lambda^{\bm{3}}_{\bm{i}} \Lambda^{\bm{3}}_{\bm{j}} \right).
\end{eqnarray}
\endnumparts
The evolution\footnote{We adopt the following notation:
  $\dot{\pi}_{\bm{rs}}=e^a_{\bm{r}} e^b_{\bm{s}}\left( u^c\nabla_c \pi_{ab} \right)$ and $\dot{q}_{\bm{r}}=e^a_{\bm{r}}\left( u^c\nabla_c q_{a} \right)$.\label{footnote-dots}} of the
thermodynamic quantities along the congruence $u^a$ can be related to the optical scalars of the congruence itself; introducing the operator
\begin{eqnarray}
\Dia \equiv AD+A^{-1}\Delta,
  \label{eq:}
\end{eqnarray}
we find
\numparts
\begin{eqnarray}
  \label{eq:raych-N-1}
  \dot{\mu}_{\mathrm{G}} &=& \frac{c}{2\sqrt{2}}\,\Dia \left( |\Psi_4| A^{-2} \right)\,, \label{eq:N-mudot}
\\[10pt]
  \dot{q}_{\bm{0}} &=& - \muG\,\dot{u}_{\bm{3}}, \\[10pt]
  \dot{q}_{\bm{1}} &=& \frac{c}{4\,\sqrt{2}}\,\frac{|\Psi_4|}{A^2} \left( \bar{\tau} + \tau + A^{-2}(\nu+\bar{\nu})\right), \\[10pt]
  \dot{q}_{\bm{2}} &=& \frac{\ii\,c}{4\sqrt{2}}\,\frac{|\Psi_4|}{A^2} \left( \bar{\tau}-\tau + A^{-2}(\nu-\bar{\nu})\right), \\[10pt]
  \dot{q}_{\bm{3}} &=& - \dot{\mu}_{\mathrm{G}}\, , \\[10pt]
 \dot{\pi}_{\bm{01}} &=& -\dfrac{1}{3}\,\dot{u}_{\bm{1}} \,, \quad
                         \dot{\pi}_{\bm{02}} = -\dfrac{1}{3}\,\dot{u}_{\bm{2}} \,, \quad
                                              \dot{\pi}_{\bm{03}}  = \dfrac{2}{3}\,\dot{u}_{\bm{3}} \,, \\[10pt]
  \dot{\pi}_{\bm{23}} &=& -\dot{q}_{\bm{2}} \,, \quad \quad
                         \dot{\pi}_{\bm{13}} = -\dot{q}_{\bm{1}} \,, \quad \quad
                                                 \dot{\pi}_{\bm{12}} = 0\, ,\\[10pt]
  \dot{\pi}_{\bm{11}} &=& \dot{\pi}_{\bm{22}}= -\frac{1}{2}\dot{\pi}_{\bm{33}} = 
  -\frac{c}{6\sqrt{2}} A^{-2}\, \Dia|\Psi_4|\,.
\end{eqnarray}
\endnumparts
Notice that the Bianchi identities cannot be used in order to eliminate $\Delta$ and $\bar{\delta}$ derivatives of $\Psi_4$ since $\Psi_4$ can be freely specified on an initial null hypersurface.  Notice that the evolution equation for $\muG$ is equivalent to the first law defined by \eqref{eq:xia-Ja}.  The force density, \ie the covariant divergence of the SQBR, is given in Type N by
\begin{eqnarray}
F^a &=& \frac{c}{2}\, |\Psi_4|\, \left( \rho + \bar{\rho} \right)\, \ell^a\,,
\label{eq:}\end{eqnarray}
and it is hence governed exclusively by the expansion of the $\ell^a$ congruence. The explicit expressions of the terms in the first law (\ref{eq:1st-law}) read
\numparts
\begin{eqnarray}
  \label{eq:N-1st-law}
  \delta Q &=& \frac{c}{2\sqrt{2}}\,|\Psi_4| \left( A^{-1}(\mu+\bar{\mu}-\gamma-\bar{\gamma}) + \frac{A}{2}(\rho+\bar{\rho}) + DA - 3\,A^{-2}\,\Delta A\right) \nonumber \\
           &&+ \frac{c}{2\sqrt{2}}\,A^{-1}\,\Delta|\Psi_4|\,, \\
  \delta W &=& \frac{c}{2\sqrt{2}}\,A^{-3}\,|\Psi_4| \left(- \gamma-\bar{\gamma} + A\,DA - A^{-1}\Delta A\right), \\
  F_a\,u^a &=& \frac{c}{2\sqrt{2}}\,A^{-1}(\rho+\bar{\rho})|\Psi_4|\,.
\end{eqnarray}
\endnumparts

\subsection{pp-wave spacetime}

An important example of Type N spacetime is that of plane-parallel waves ({\itshape pp-waves}) \cite{Stewart-1993}. In the  coordinates $x^\mu=(u,\, v,\, \zeta,\, \bar{\zeta})$, where
$\zeta = x + \ii\,y$, the line element reads
\begin{eqnarray}
  \label{eq:pp-wave-metric}
  \dd s^2 = 2\,\dd u\,\dd v + 2\,H(u,\zeta,\bar{\zeta})\, du^2 - 2 \,\dd \zeta\,\dd\bar{\zeta}\ ,
\end{eqnarray}
where $H$ is an arbitrary function harmonic in the coordinates $x, y$.  A convenient choice of the null tetrad is
\begin{equation}
\ell  = \pd_v\ , \qquad
n    = \pd_u-H(u,\,\zeta,\,\bar{\zeta})\, \pd_v\ , \qquad
m    = \pd_\zeta\, ,
\label{eq:}\end{equation}
in which the only non-vanishing NP scalars are
\begin{equation}
\Psi_4  = H_{,\bar{\zeta}\bar{\zeta}}\ , \qquad
\nu  = H_{,\bar{\zeta}}\, ,
\label{eq:}\end{equation}
where comma means derivative with respect to the indicated variables.  Thus, using $\Delta|\Psi_4| = |\Psi_4|_{,u}$, the first law and the evolution of the heat flux read
\numparts
\begin{eqnarray}
\dot{\mu}_{\mathrm{G}} &=& 
  -\frac{c}{2\sqrt{2}} \frac{1}{A^4}\left\{ 2|\Psi_4|\left[ \left(A^2-H\right)\,A_{,v}+A_{,u} \right] - A|\Psi_4|_{,u} \right\} , \\
\dot{q}_{\bm{0}} &=& \frac{1}{2\sqrt{2}}\,\frac{|\Psi_4|_{,u}}{A^3} - \frac{\Grav{\dot{\mu}}}{2}\,, \\
\dot{q}_{\bm{1}} &=& \frac{1}{4\sqrt{2}}\, \frac{|\Psi_4|}{A^4} \left( H_{,\zeta}+H_{,\bar{\zeta}} \right) , \\
\dot{q}_{\bm{2}} &=& \frac{\ii}{4\sqrt{2}}\, \frac{|\Psi_4|}{A^4} \left( H_{,\zeta}-H_{,\bar{\zeta}}\right) , \\
\dot{q}_{\bm{3}} &=& -\Grav{\dot{\mu}} \,.
\end{eqnarray}
\endnumparts
The expansion of the congruence (\ref{eq:ua-boost}) is
\begin{eqnarray}
  \label{eq:N-expansion}
  \theta = \frac{1}{\sqrt{2}\,A^2}\left[ \left( A^2 + H\right)A_{,v} - A_{,u} \right],
\end{eqnarray}
so that the variation of the volume reads
\begin{eqnarray}
  \label{eq:N-deltaV}
 \left( \delta V_{\mu\nu} \right) \dd x^\mu\,\dd x^\nu &=&
  - \frac{\theta}{2\,A^2}\left( \dd v^2 + (A^2-H)^2 \dd u^2 \right) + \theta\left(1 - \frac{H}{A^2}\right)\,\dd u\,\dd v \nonumber \\
  && + \frac{1}{2\,\sqrt{2}\,A^3}\left( H_{,\zeta}\,\dd \zeta + H_{,\bar{\zeta}}\,\dd\bar{\zeta}\right)\left(\dd v - (A^2-H)\,\dd u\right).
\end{eqnarray}
We notice that in this case the SQBR is fully conserved, in the sense that $F^a \equiv 0$, as the expansion of the $\ell^a$ congruence vanishes. Consequently, the terms $\delta W$ and $\delta Q$ are the only ones contributing to the first law (\ref{eq:1st-law}):
\begin{eqnarray}
  \label{eq:N-deltaW-deltaQ}
  \delta W &=& \frac{c}{2\,A^2}\,\theta\,|\Psi_4|\,, \\
  \delta Q &=& \frac{c}{2\sqrt{2} \,A^4}\left( A\,|\Psi_4|_{,u}  + 4\,|\Psi_4|(H\,A_{,v}-A_{,u})\right)\,.
\end{eqnarray}
Notice that, for the choice
\begin{eqnarray}
  \label{eq:N-A}
  A &=& A_0(\zeta, \bar{\zeta})\,\sqrt{|\Psi_4|}\,,
\end{eqnarray}
where $A_0$ is an arbitrary function of the arguments indicated, \eqref{eq:N-mudot} implies that $\Grav{\dot{\mu}}=0$, which means that gravitational energy is conserved along such congruence.  Then, the expansion takes the form
\begin{eqnarray}
  \label{eq:N-expansion}
  \theta &=& \frac{1}{\sqrt{2}\,A_0}\,(|\Psi_4|^{-1/2})_{,u}\,,
\end{eqnarray}
and the work and heat terms take the form
\begin{eqnarray}
  \label{eq:N-deltaQ-W}
  \delta Q &=& 2\,\delta W = c\,\theta\,A_0^{-2}.
\end{eqnarray}

\section{Thermodynamics of Type D spacetimes}
\label{sec:thermo-D}

In Type D spacetimes, with the same choice of frame given by \eqref{eq:ua-boost}, the electromagnetic spinor acquires the form \eqref{eq:D-phiAB-basis} and, in term of the fields, $\phi_1$ reads
\begin{eqnarray}
 \phi_1 = - E_z -\ii\, B_z = \sqrt{\Psi_2}\, .
\end{eqnarray}
This implies that $E^a$ and $B^a$ are parallel, so that for such observers $q_a=0$.  However, transforming to a general frame, $q_a$ becomes non-vanishing, which is analogous to the presence
of dissipative effects for non-comoving observers in a fluid. The reason is that, contrary to Type N, in this case $q_a$ does not represent a true flux of gravitational energy and has purely kinematical origin.  In the frame adapted to the principal null directions, which we henceforth call \emph{comoving}, the thermodynamic quantities in terms of $\Psi_2$ read
\numparts
\begin{eqnarray}
  \label{eq:thermo-D}
  \muG &=& c|\Psi_2|\, , \\
  \PG &=& - \frac{c}{3}|\Psi_2|\, , \\
  q_a &=& 0\, , \\
  \pi_{ab} &=& \frac{2\, c}{3} |\Psi_2| \left( x_a\,x_b+y_a\,y_b - 2\,z_a\,z_b\right).
\end{eqnarray}
\endnumparts
Transforming the comoving frame to a generic one as explained in \ref{app:lorentz}, we get
\numparts
\begin{eqnarray}
  \label{eq:D-lorentz}
  \Grav{\tilde{\mu}} &=& (\muG+\PG) (\Lambda^{\bm{0}}_{\bm{0}})^2 - (\PG-\pi_{\bm{11}}) \sum_{\bm{i}=1}^3 (\Lambda^{\bm{i}}_{\bm{0}})^2 - 3\,\pi_{\bm{11}}\,(\Lambda_{\bm{0}}^{\bm{3}})^2, \\
  \Grav{\tilde{P}} &=& - \frac{1}{3}\,\Grav{\tilde{\mu}}, \label{eosD} \\
  \tilde{q}_{\bm{i}} &=& (\muG+\PG)\Lambda^{\bm{0}}_{\bm{i}}\,\Lambda^{\bm{0}}_{\bm{0}} - (\PG-\pi_{\bm{11}})\sum_{\bm{j}=1}^3 \Lambda^{\bm{j}}_{\bm{i}}\,\Lambda^{\bm{j}}_{\bm{0}} -
                         3\,\pi_{\bm{11}}\,\Lambda^{\bm{3}}_{\bm{i}}\,\Lambda^{\bm{3}}_{\bm{0}}\,, \\
  \tilde{\pi}_{\bm{ij}} &=& (\muG+\PG)\Lambda^{\bm{0}}_{\bm{i}}\,\Lambda^{\bm{0}}_{\bm{j}} - (\PG-\pi_{\bm{11}})\sum_{\bm{k}=1}^3 \Lambda^{\bm{k}}_{\bm{i}}\,\Lambda^{\bm{k}}_{\bm{j}} -
                         3\,\pi_{\bm{11}}\,\Lambda^{\bm{3}}_{\bm{i}}\,\Lambda^{\bm{3}}_{\bm{j}} - \Grav{\tilde{P}}\,\delta_{\bm{ij}}\,.
\end{eqnarray}

\endnumparts
The evolution of such quantities along the timelike congruence $u^a$ is given by
\numparts
\begin{eqnarray}
  \label{eq:raych-D}
  \Grav{\dot{\mu}} + \Grav{\dot{P}} &=& \frac{3}{2\sqrt{2}}\, \left(\muG + \PG \right)\, \left[ A\left( \rho+\bar{\rho} \right)- A^{-1}\left( \mu+\bar{\mu} \right) \right]\,,
\end{eqnarray}
\begin{eqnarray}
  \dot{\pi}_{\bm{01}} &= \left( \muG+\PG \right) \dot{u}_{\bm{1}}\,, \quad
  &\dot{\pi}_{\bm{02}} = \left( \muG+\PG \right) \dot{u}_{\bm{2}}\,, \\
  \dot{\pi}_{\bm{03}} &= -2\left( \muG+\PG \right) \dot{u}_{\bm{3}}\,, 
\end{eqnarray}
\begin{equation}
  \dot{\pi}_{\bm{11}} = 
  \dot{\pi}_{\bm{22}} = 
 -\frac{1}{2}\dot{\pi}_{\bm{33}} = \Grav{\dot{\mu}}+\Grav{\dot{P}}\,,
\end{equation}
\begin{eqnarray}
  \dot{\pi}_{\bm{12}} &=& 0\, ,\\
\dot{\pi}_{\bm{13}} &=& \frac{3}{2\sqrt{2}}\, \left(\muG+\PG\right)\, \left[ \tau + \bar{\tau} + \pi + \bar{\pi} \right] ,\\
\dot{\pi}_{\bm{23}} &=& \frac{3\, \ii}{2\sqrt{2}}\, \left( \muG+\PG \right)\, \left[ \tau - \bar{\tau} -\pi +\bar{\pi} \right]\,.
\end{eqnarray}
\endnumparts
The covariant divergence of the SQBR yields the force density
\begin{eqnarray}
F^a &=&- \frac{c}{2}\, |\Psi_2|\, \Big[ \left( \mu+\bar{\mu} \right)\ell^a- \left( \rho+\bar{\rho} \right)n^a -\left( \bar{\tau}-\pi \right)m^a -\left( \tau-\bar{\pi} \right)\bar{m}^a\Big]\,.
\label{eq:}\end{eqnarray}
With regard to the first law, since the heat flux $q^a$ vanishes for the comoving congruence, the dissipative term is identically zero, \ie $\delta Q = 0$.  Moreover, it is straightforward to show
that
\begin{eqnarray}
  \label{eq:D-deltaW}
\fl  \delta W = \frac{c}{\sqrt{2}}|\Psi_2|\left[ - A (\eps+\bar{\eps}+\rho+\bar{\rho}) + A^{-1}(\gamma+\bar{\gamma}+\mu+\bar{\mu})- DA +A^{-2}\Delta A\right]\,,\\
\fl  F^au_a = \frac{c}{2\sqrt{2}}\, |\Psi_2|\, \left[ A\left(\rho+\bar{\rho}\right) -A^{-1}\left( \mu+\bar{\mu} \right)\right]\,,
\end{eqnarray}
so that the first law for observers $u^a$ can be written as
\begin{eqnarray}
 \label{eq:1st-law-D}
 \Grav{\dot{\mu}} = -\muG\, \left(\theta - 3\, \sigma_{\bm{11}} \right)= -\frac{3}{2\sqrt{2}}\, \muG \, \left[ -A\left( \rho+\bar{\rho} \right)+ A^{-1}\left( \mu+\bar{\mu} \right) \right]\,,
\end{eqnarray}
where we have recast the equation also in terms of optical scalars and we have used the redundancy between $\PG$ and $\muG$ to express the law exclusively in terms of the latter.  However notice that this is formally the same as \eqref{eq:raych-D}.  Energy is conserved along the timelike congruence if $\Grav{\dot{\mu}}=0$ and this happens whenever $\theta=3\, \sigma_{\bm{11}}$ is satisfied.  We notice that the condition for energy conservation brings as well to $F^au_a=0$.  Below we specialize to the case of Kerr black hole metric and find the specific observers with this property.

\subsection{Kerr black hole and Carter observers}
In the Boyer-Lindquist coordinates, the null tetrad adapted to principal null directions reads
\numparts
\begin{eqnarray}
  \ell &=& \frac{r^2+a^2}{\KDelta}\,\pd_t + \pd_r  + \frac{a}{\KDelta}\,\pd_\phi, \\
  n &=& \frac{1}{2\Sigma}\left( (r^2+a^2)\,\pd_t  - \KDelta\,\pd_r+a\,\pd_\phi\right),  \\
  m &=& \frac{1}{\sqrt{2}\,\Gamma}\left( \ii \,a \sin\theta\, \pd_t  + \pd_\theta + \frac{\ii}{\sin\theta}\,\pd_\phi\right),
        \label{eq:Kerr-null-tetrad}
\end{eqnarray}
\endnumparts
where
\begin{eqnarray}
  \label{eq:Kerr-definitions}
  \KDelta = r^2  - 2\,M\,r + a^2, \qquad \Gamma = r + \ii\,a\cos\theta\,, \qquad \Sigma=r^2+a^2\cos^2\theta\,.
\end{eqnarray}
Being Type D, Kerr spacetime has only one non-vanishing Weyl scalar, namely
\begin{eqnarray}
  \label{eq:Kerr-Psi2}
  \Psi_2 &= - \frac{M}{\bar{\Gamma}^3},
\end{eqnarray}
so that the gravitational energy density is given by
\begin{eqnarray}
  \label{eq:Kerr-muG}
  \muG &=& \frac{c\,M}{\Sigma^{3/2}}\,.
\end{eqnarray}
The non-vanishing spin coefficients read
\numparts
\begin{eqnarray}
  \label{eq:Kerr-spins}
  \mu &=-\frac{\KDelta}{2\, \Sigma\, \bar{\Gamma}}\,, \qquad &\gamma = \mu+\frac{r-M}{2\Sigma}, \\
  \pi &= \frac{\ii\,a\sin\theta}{\sqrt{2}\,\bar{\Gamma}^2}\,, &\alpha = \pi-\bar{\beta} \,, \qquad \rho = - \frac{1}{\bar{\Gamma}}\,,\\
  \tau &= - \frac{\ii\,a\sin\theta}{\sqrt{2}\,\Sigma}\, , & \beta = \frac{\cot \theta}{2\sqrt{2}\,\Gamma}\,.
\end{eqnarray}
\endnumparts
The first law \eqref{eq:1st-law-D} now takes the form
\begin{equation}
\Grav{\dot{\mu}} = - \frac{3}{2\sqrt{2}}\frac{r}{\Sigma} \left[ 2A-\frac{\KDelta}{A\,\Sigma} \right]\, \muG\, .
\end{equation}
It is instructive to show all the components of the first law as expressed in \eqref{eq:1st-law}:
\numparts
\begin{eqnarray}
\fl \delta Q &=& 0\,, \\
\fl  \delta W &=& \frac{c\,M}{\sqrt{2}\,\Sigma^{5/2}}\left[  2\,A\,r+A^{-1}\left(r-M -2\,r\,\KDelta\,\Sigma^{-1}\right) -\Sigma\left(DA -A^{-2}\,\Delta A\right) \right]\,, \\
\fl  F_au^a &=& \frac{c\, M}{2\sqrt{2}}\, \frac{r}{ \Sigma^{5/2}}\, \left(\frac{\KDelta}{A\, \Sigma} - 2\,A \right)\,.
\end{eqnarray}
\endnumparts
There exists a class of observers for which $F^au_a=\delta W=0$ simultaneously, corresponding to the choice
\begin{equation}
A = \Cart{A} \equiv \sqrt{\frac{\KDelta}{2\, \Sigma}}\, .
\label{eq:Acarter}
\end{equation}
Such form of the boost parameter identifies the so-called \emph{Carter observers}, for which
\begin{equation}
\Cart{u} = \frac{1}{\sqrt{\Sigma\, \KDelta}} \left[ \left( r^2+a^2 \right)\pd_t+a\, \pd_\phi \right]\,.
\label{eq:}\end{equation}
These trajectories are the only ones whose 4-velocities belong to the intersection of the $\ell-n$ plane with the Killing plane $t-\phi$.  Apart from possessing the symmetries that allow exact integration of the geodesic equation, it has been noticed in the literature \cite{bini2012tidal} that for such observers the \emph{super-Poynting vector} vanishes and the \emph{super-energy} is minimized.  In fact, the vanishing of the super-Poynting vector is a feature of all congruences with generic boost in the $\ell-n$ plane.  We have shown here, further, that the \emph{heat flux} $q_a$ vanishes for a class of observers broader than the Carter ones (in fact, for any choice of $A$ in the adapted tetrad) and that the \emph{energy} $\muG$ defined by the SQBR is conserved along $\Cart{u}$. Notice that, being the force density, the quantity $F_a$ is a purely spatial vector and Carter frame is the one for which its timelike component vanishes. Nevertheless, spatial components of $F_a$ are non-zero and read
\begin{eqnarray}
  \label{eq:D-F-comps}
\fl  F_{\bm{1}} = - \frac{c\,a^2\,r\,M \cos\theta \sin\theta}{\Sigma^{7/2}}\,, \quad
  F_{\bm{2}} =  \frac{c\,a^3\,M \cos^2\theta \sin\theta}{\Sigma^{7/2}}\,, \quad
  F_{\bm{3}} = - \frac{c\,r\,M\,\sqrt{\KDelta}}{\Sigma^3}\,.
\end{eqnarray}
We stress again that $F_a$ is an intrinsic quantity and, as such, it cannot be gauged away. Finally, it is easy to check that in the limit of non-rotating
black hole spacetime, \ie $a\to 0$, the congruence $\Cart{u}$ corresponds to static observers at fixed radial distance.

\section{Final remarks}\label{sec5:concl}

In the present paper we have provided a detailed analysis of the properties of the SQBR interpreted as an energy-momentum tensor for the gravitational field.  In analogy with the behavior of any massless field theory, the tracelessness of the EMT has been chosen as a fundamental constraint on the definition of the SQBR.  Further, we have provided both a fluid-like and an electromagnetic-like description of the SQBR, finding explicit relations between optical scalars of the timelike congruences associated with observers and the thermodynamic and electromagnetic quantities associated with the spacetime geometry.  We conclude that the gravitational field, as described by the SQBR under the aforementioned constraint, is a genuinely dissipative system because the energy-momentum tensor is not covariantly conserved.  In fact, the deviation from conservation as expressed by the balance equation \eqref{eq:Fa-def} is not an observer-dependent effect and hence cannot in general be attributed to a choice of frame.  One could conjecture that such intrinsic dissipation could be related to a transfer of energy between the macroscopic gravitational field and its underlying microscopic degrees of freedom, in the spirit of \cite{Acquaviva2017,chirco2010nonequilibrium}.  Nevertheless, pp-wave spacetime is an example where this intrinsic dissipation vanishes identically.

We have proposed a generalized first law of gravitational thermodynamics arising from the timelike component of the balance equation, \ie $\delta U = \delta Q - \delta W - F_au^a$.  Such first law contains terms which can be interpreted as due to work and to dissipation.  The work term $\delta W$ arises from the deformation of the volume element along the observer's wordline.  Among the dissipative terms, $\delta Q$ is directly related to the heat flux $q^a$ and represents the transfer of energy between parts of the system.  In Type N spacetimes, which describe propagation of gravitational waves, this term cannot be killed by a choice of frame; in Type D, instead, there exist \emph{comoving frames} in which $q^a$ and hence $\delta Q$ are zero.  The dissipative term $F_au^a$, being only the timelike component of the intrinsic dissipation, can be gauged away for specific choices of the observer.  

An important ingredient in the description of ordinary fluids is the equation of state, relating the pressure to the energy density.  The role of the equation of state here is played by the constraint imposed on the free function $f$ in \eqref{eq:SQBR-freedom}: with our choice $f=0$, which makes the SQBR traceless (cf.\ Sec.\ \ref{sec:constraints}), we obtain $\PG=-1/3\, \muG$ which corresponds in our conventions to a radiation-like fluid.  This is true for both Type N and D spacetimes.  Moreover, unlike other observer-dependent effects, the form of the equation of state is invariant under general Lorentz transformations (see \eqref{eosN} and \eqref{eosD}).

We exemplified such results in the case of Type N for a pp-wave metric.  We have shown that the SQBR in these spacetimes is covariantly conserved, \ie $F^a\equiv0$ and that
there exists a congruence of observers for which the energy density is conserved. The only contribution to the change of the internal energy $\delta U$ in this case
is solely due to the expansion of the congruence itself.  In Type D spacetimes, we have analyzed the thermodynamic properties of the gravitational field in the specific case of Kerr black holes in comoving frames, for which $\delta Q=0$.  Among such comoving frames, a particular subclass known in the literature as \emph{Carter observers} have in addition the property of conservation of gravitational energy density, \ie $\Cart{u}^a\nabla_a\, \muG=0$.  We stress the fact that such property is not due to a fortuitous cancellation between the various terms in the first law: in fact, apart from having vanishing expansion $\theta_u$, for Carter observers the work term $\delta W$ and the dissipation $F_au^a$ are separately identically zero.

In the present paper we assumed that the SQBR provides a correct thermodynamic description of the gravitational field and hence investigated the interpretation of the results directly following from this assumption.  However, the physical viability of this approach is still matter of research.  In particular, in order to justify the definition of gravitational energy based on SQBR one should investigate its limits in canonical examples, like ADM mass or linearized gravity.  Moreover, in this paper we formulated the first law \emph{locally} in terms of densities, while for a full thermodynamic description one would need a \emph{quasi-local} formulation of thermodynamic laws and a proper definition of variations of quasi-local quantites.  This would also help in clarifying the role of the dissipative effects encountered in the present analysis.  These issues will be addressed in subsequent works.

\ack

This work was supported by the GA\v{C}R grant 17-16260Y of the Czech Science Foundation.  GA and MS thank L\'{a}szl\'{o} Szabados and Rituparno Goswami for the kind hospitality and useful discussions.  We thank the anonymous referees for enlightening comments that lead to improvements in the paper.

\appendix

\section{Lorentz transformations}
\label{app:lorentz}

Consider the orthonormal tetrad (\ref{eq:ua-boost})--(\ref{eq:tetradz}) induced by a null tetrad adapted to principal null directions of given Type N or Type D spacetime. We interpret
such tetrad as the one associated with a comoving observer for whom the thermodynamic quantities acquire their simplest form. There are several reasons why we wish to
consider a general observer, though. First, $u^a$ given by (\ref{eq:ua-boost}) is in general not hypersurface orthogonal and hence does not define a foliation of the spacetime
by spacelike hypersurfaces. Such foliation is necessary if one wants to introduce quasi-local thermodynamic quantities as integrals over spatial domains. Second, in accordance
with the fluid interpretation presented in Sec.\ \ref{sec:fluid-interpretation}, some effects like the presence of a heat flux are observer-dependent.

In order to discuss generic observers, we consider a general Lorentz transformation of the comoving tetrad induced by a sequence of basic types of transformations of the
spin basis \cite{Penrose-Rindler-I}. The most general transformation of the spin basis reads
\begin{eqnarray}
  \label{eq:lorentz-transf}
  \tilde{o}^A = L\,e^{\ii\,\chi}\,o^A + R\,\iota^A, \qquad \tilde{\iota}^A = L^{-1}\,e^{-\ii\,\chi}\,\iota^A + S\,\tilde{o}^A,
\end{eqnarray}
where the real parameter $L$ represents the boost in the plane spanned by $\ell^a$ and $n^a$, the real parameter $\chi$ induces rotation in the plane
spanned by $m^a$ and $\bar{m}^a$, and the complex parameters $R$ and $S$ represent null rotations with fixed $\ell^a$ and $n^a$, respectively. The transformation of the spin basis
gives rise to the Lorentz transformation
\begin{eqnarray}
  \label{eq:lorentz-tetrad}
  \tilde{e}^a_{\bm{a}} = \Lambda^a_b\,e^b_{\bm{a}}\,,
\end{eqnarray}
where the tetrad components of $\Lambda^a_b$ are given by $\Lambda^{\bm{b}}_{\bm{a}} = \tilde{e}^a_{\bm{a}}\,e^{\bm{b}}_a$.  Explicitly we have
\begin{eqnarray}
  \label{eq:Lambda-matrix-1}
  \Lambda^{\bm{0}}_{\bm{0}} &=& \frac{1}{2}(L^{2}+L^{-2}) + \frac{1}{L}\,\Re( e^{\ii\,\chi} R\,S) + \frac{1}{2}(|R|^2 + L^2\,|S|^2+| R\,S|^2), \\
  \Lambda^{\bm{1}}_{\bm{0}} &=&  \Re\left( e^{\ii\,\chi}\,L\,\bar{R}(1+|S|^2)+S\,e^{2\,\ii\,\chi}\right),\\
  \Lambda^{\bm{2}}_{\bm{0}} &=& - \Im\left( L\,\bar{R}\,e^{\ii\,\chi}(1+|S|^2) + S\,e^{2\,\ii\,\chi}\right), \\
  \Lambda^{\bm{3}}_{\bm{0}} &=&  -\frac{1}{2}(L^{-2}-L^2) - \frac{1}{L} \Re( e^{\ii\,\chi}\,R\,S) - \frac{1}{2}(|R|^2-L^2\,|S|^2+|R\,S|^2),
\end{eqnarray}
\begin{eqnarray}
  \label{eq:Lambda-matrix-2}
  \Lambda^{\bm{0}}_{\bm{1}} &=& \frac{1}{L} \Re(R\,e^{\ii\,\chi}) + (L^2 + |R|^2)\Re S, \\
  \Lambda^{\bm{1}}_{\bm{1}} &=&  \cos 2\chi + 2\,L\,(\Re S)\,\Re( \bar{R}\,e^{\ii\,\chi}), \\
  \Lambda^{\bm{2}}_{\bm{1}} &=& - \sin 2\chi - 2\,L(\Re S)\,\Im(\bar{R}\,e^{\ii\,\chi}), \\
  \Lambda^{\bm{3}}_{\bm{1}} &=& -\frac{1}{L}\,\Re( R\,e^{\ii\,\chi}) - (|R|^2 - L^2)\Re S,
\end{eqnarray}
\begin{eqnarray}
  \label{eq:Lambda-matrix-3}
  \Lambda^{\bm{0}}_{\bm{2}} &=& \frac{1}{L}\,\Im (R\,e^{\ii\,\chi}) - (L^2 + |R|^2)\,\Im S, \\
  \Lambda^{\bm{1}}_{\bm{2}} &=&  \sin 2\chi - 2\,L(\Im S)\,\Re(\bar{R}\,e^{\ii\,\chi}), \\
  \Lambda^{\bm{2}}_{\bm{2}} &=&  \cos 2\chi + 2\,L(\Im S)\,\Im( \bar{R}\,e^{\ii\,\chi}), \\
  \Lambda^{\bm{3}}_{\bm{2}} &=& -\frac{1}{L}\,\Im( R\,e^{\ii\,\chi}) - (L^2 - |R|^2)\Im S, 
\end{eqnarray}
\begin{eqnarray}
  \label{eq:Lambda-matrix-4}
  \Lambda^{\bm{0}}_{\bm{3}} &=& \frac{1}{2}(L^2 + L^{-2}) - \frac{1}{L}\,\Re(R\,S\,e^{\ii\,\chi}) + \frac{1}{2}(|R|^2-L^2\,|S|^2-|R\,S|^2), \\
  \Lambda^{\bm{1}}_{\bm{3}} &=& -\Re\left( S\,e^{2\,\ii\,\chi} + L(1+|S|^2)\,\bar{R}\,e^{\ii\,\chi} \right),\\
  \Lambda^{\bm{2}}_{\bm{3}} &=&- L(1+|S|^2)\,\Im ( \bar{R}\,e^{\ii\,\chi}) + \Im ( S\,e^{2\,\ii\,\chi} ), \\
  \Lambda^{\bm{3}}_{\bm{3}} &=&  \frac{1}{2}(L^2 + L^{-2}) + \frac{1}{L} \,\Re( R\,S\,e^{\ii\,\chi}) - \frac{1}{2} \left( |R|^2 + L^2\,|S|^2 - |R\,S|^2\right),
\end{eqnarray}
where $\Re$ and $\Im$ represent the real and the imaginary parts, respectively.

\section*{References}

\providecommand{\newblock}{}

\end{document}